\documentclass[pdftex,final,1p,times,twocolumn]{elsarticle}

\pdfoutput=1
\usepackage{graphics}

\usepackage{amssymb}

\biboptions{square, comma, sort&compress}

\journal{High Energy Density Physics}

\begin{document}

\begin{frontmatter}



\title{Irradiated Interfaces in the Ara OB1, Carina, Eagle Nebula, and Cyg OB2 Massive Star Formation Regions}

\author[1,2]{P. Hartigan}
\author[1,2]{J. Palmer}
\author[2,3]{L.I. Cleeves}
\address[1]{Physics and Astronomy Dept., Rice University, Houston TX}
\address[2]{Visiting Astronomer, NOAO Observatories}
\address[3]{Astronomy Dept., University of Michigan, Ann Arbor MI}

\begin{abstract}
Regions of massive star formation offer some of the best and most
easily-observed examples of radiation hydrodynamics. Boundaries where 
fully-ionized H II regions transition to neutral/molecular photodissociation
regions (PDRs) are of particular interest because marked temperature and density
contrasts across the boundaries lead to evaporative flows and fluid dynamical
instabilities that can evolve into spectacular pillar-like structures.  
When detached from their parent clouds, pillars become ionized globules 
that often harbor one or more young stars. 
H$_2$ molecules at the interface between a PDR and an H~II region absorb ultraviolet light
from massive stars, and the resulting fluoresced infrared emission lines are an
ideal way to trace this boundary independent of obscuring dust.  This paper presents H$_2$ images of four
regions of massive star formation that illustrate different types of PDR boundaries.
The Ara OB1 star formation region contains a striking long wall that has several wavy structures which 
are present in H$_2$, but the emission is not particularly bright because the ambient
UV fluxes are relatively low.  In contrast, the Carina star formation region shows strong H$_2$ fluorescence 
both along curved walls and at the edges of spectacular pillars
that in some cases have become detached from their parent clouds. The less-spectacular
but more well-known Eagle Nebula has two regions that have strong fluorescence in addition to
its pillars.  While somewhat older than the other regions,
Cyg OB2 has the highest number of massive stars of the regions surveyed and contains
many isolated, fluoresced globules that have head-tail morphologies which point
towards the sources of ionizing radiation.  These images provide a collection of
potential astrophysical analogs that may relate to ablated interfaces observed in
laser experiments of radiation hydrodynamics.
\end{abstract}


\end{frontmatter}


\section{Introduction}

Star formation is a complex dynamical process where the attractive force of gravity must overcome 
the dispersive effects of pressure and angular momentum. Once stars form,
they provide feedback into the nascent molecular clouds by generating powerful bipolar
outflows that evacuate cavities and deposit momentum into the clouds
(see, e.g. \cite{starform1}).
In regions of star formation where the most massive star has a spectral
type earlier than $\sim$ B3 (M $\gtrsim$ 10 M$_\odot$), radiation is the primary feedback mechanism.
In these cases, intense ultraviolet radiation from the massive stars ionizes an `H II region'
within which H is nearly fully-ionized. At the boundaries of an H II region,
H quickly transitions to a high-density neutral state called a photodissociation region
or PDR. Radiation that is less-energetic than the ionization potential of H (13.6 eV) penetrates
into the PDR, and affects molecular chemistry by dissociating molecules, heating dust
grains, and ionizing elements such as C \cite{pdrref}.

Irradiated PDR interfaces are important to understand because it is here where radiation deposits energy
into and ultimately disrupts the molecular cloud.  The disruption occurs as neutral material at the
front becomes ionized and, finding itself at a high temperature, ablates into the less-dense H II
region.  The ablation flow drives pressure waves and weak shock fronts back
into the molecular cloud, heating the gas and dust near the interface and thereby
enhancing the formation rate of complex molecules in the PDR. The ablation process may also affect
star formation because gas located at the tips of pillars compresses in response to the
radiation pressure. Some studies suggest this inward pressure may trigger stars to form at
these locations, though it can be difficult to distinguish sequential star formation,
where there are simply spatial gradients in the apparent ages, from triggering
 \cite{trigger}.  The morphologies of the interfaces provide unique opportunities to
study how fluid dynamical instabilities form and develop in astrophysical systems. 

Hence, along with gravity and rotation, radiation hydrodynamics is one of the dominant
physical processes in regions of massive star formation, and one 
that is amenable to study in the laboratory.
The intense radiation environments present in laser experiments often produce
ablation flows at locations where the radiation interacts with a surface
\cite{laseref1}. As material ablates, the interfaces can become irregular
in shape and eventually transform into pillar-like structures \cite{laseref2,laseref3}
whose morphologies resemble those present in molecular clouds subject to
radiation from nearby massive stars \cite{laseref4}.

H II regions and molecular clouds are distinct phases of the interstellar medium,
and the clouds often show sharp boundaries that are easily-observed at optical wavelengths
because dust within molecular clouds absorbs optical light from background sources
such as stars or emission from the hot gas in the H II region. While these boundaries often
exhibit beautiful and complex morphologies, connecting the observed morphologies with
radiation-driven instabilities is not always a straightforward task. Depending on the
situation, other physical processes such as bulk motions of molecular clouds relative to the
gas within an H II region, internal magnetic and thermal pressure gradients, and the presence
of dense clumps within clouds may be more important than radiation in determining
how an interface appears and how it evolves with time.

To identify an interface as a promising analog for radiative-driven instabilites we need
to separate regions where the ultraviolet flux is highest from others where the radiative
flux is lower.  From an observational standpoint, it is challenging to determine exactly where the
ultraviolet radiation deposits its energy, because regions of star formation are
so dusty that lines of sight to PDR interfaces are typically opaque at both optical
and ultraviolet wavelengths. Hence, some of the apparent boundaries may simply result from a dense 
dusty cloud in the foreground that obscures a bright background H~II region. 
Fortunately, because fluorescent ultraviolet excitation of H$_2$ is always accompanied by
specific infrared emission lines that radiate as the molecule decays to its ground state,
one can image true PDR interfaces directly in these infrared lines where obscuration by dust is negligible. 

As a broad cut, one can sort massive star-forming regions by the number of O stars, which, along
with the rare luminous blue variable (LBV) and Wolf-Rayet (WR) stars,
represent the classes of the most massive stars. The O-spectral class is subdivided
by increasing mass and photospheric temperature into O9, O8, O7, and so on, up to about O3 and O2,
which are the most massive O stars known. Less-massive stars
are always more common, so, for example, a region that contains an O3 star will typically have many O5 $-$ O9 
stars.  The most massive stars should become supernovae within a few million years of their birth, and
undoubtedly influence the ongoing star formation in their vicinity, which would otherwise last for tens
of millions of years. While they are present, the UV radiation from the most massive stars will
dominate the ionization of its surroundings because the most massive stars have the
hottest photospheres and therefore the highest photon luminosities $>$ 13.6 eV. The UV luminosity (photons/second)
depends strongly upon the photospheric temperature because the UV part of stellar spectra lies
on the Wien portion of the blackbody curve.  For example, a typical O3 star emits an order of magnitude more
photons capable of ionizing H than does an O7.5 star, and nearly two orders of magnitude more than an
O9.5 star \cite{carinacensus}.

This paper presents near-IR images of the PDR interfaces of four regions of massive star formation
that span a range of ultraviolet radiation fields and 
morphologies, from large wall-like structures in a region with only a few O stars (Ara OB1), to spectacular pillars
within more massive regions (Carina and Eagle), to isolated globules in an extremely massive and somewhat older complex
(Cygnus OB2).  The goal is to provide the laboratory astrophysics community with several examples of typical
ablation interfaces that are present within regions of massive star formation,
as these may prove to be useful analogs for ongoing
or planned experimental work. The next section describes the physics of H$_2$ fluorescence along with the
observational program, and the paper continues with a description of what the new images show
in each of the four regions. 

\section{Infrared Observations of Fluorescent H$_2$}

Figure~1 summarizes the physics of fluorescent excitation of H$_2$.
Ultraviolet photons with energies less than the ionization potential of H (13.6 eV)
penetrate into the region of neutral gas where molecular hydrogen exists.
If their energy is greater than about 10 eV, these photons can excite
H$_2$ from its ground electronic state to the Lyman and Werner bands.
These bands can be visualized as resulting from 
the overlap of the n=2 levels of the indivdual H atoms atoms
when the protons are separated by the intranuclear distance of $\sim$ 1\AA \cite{h2ref}. 
As excited H$_2$ molecules cascade back to their ground states, there is a
probability that they will pass through the J = 3 (rotational) v = 1 (vibrational) level.
Because H$_2$ is a symmetric molecule, it has no dipole moment, and transitions of
$\Delta$J = $\pm$1 do not occur. However, the molecule can decay via $\Delta$J = $-$2
quadrupole radiation to the J = 1 v = 0 level by emitting an infrared photon at 2.12$\mu$m. 
Designated as the 1-0S(1) transition (S indicating $\Delta$J = $-$2 and `1' referring to
the J-value of the lower state), this emission line traces the interface where the
ultraviolet radiation encounters molecular hydrogen.

\begin{figure}[b]
\includegraphics{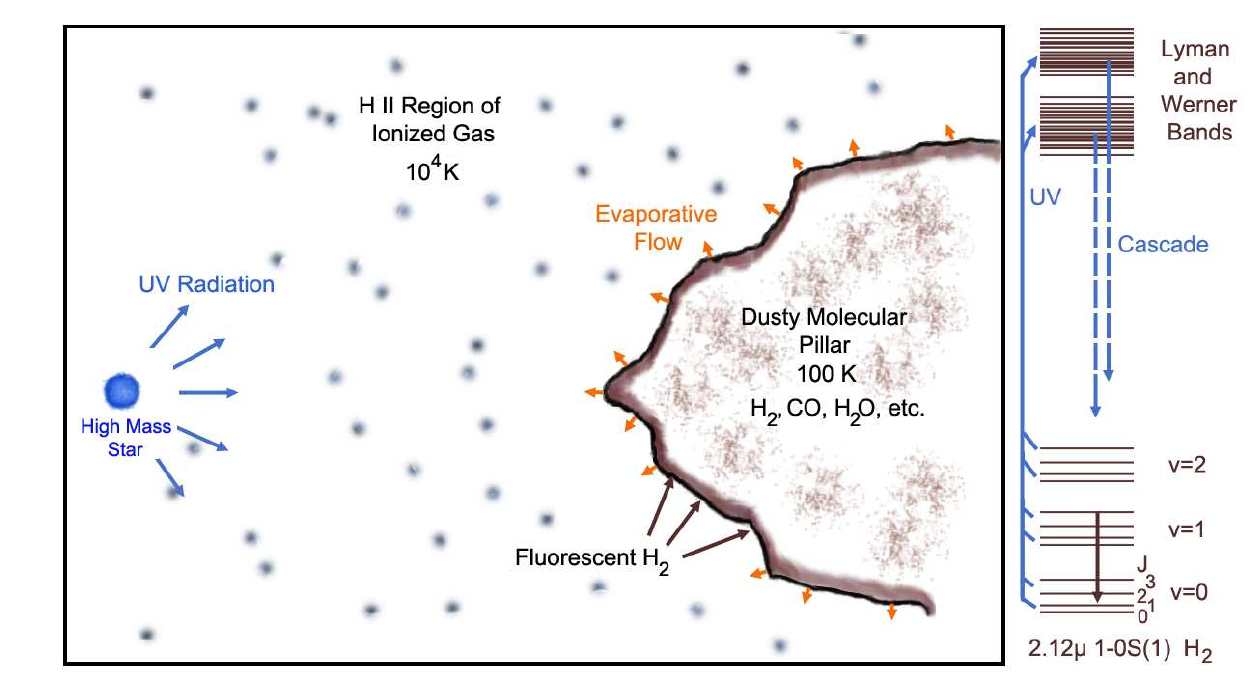}
\caption{Schematic of an H II/photodissociation region and its fluorescent H$_2$. 
Ultraviolet light from the massive star excites H$_2$ to the Lyman and Werner bands.
The 2.12$\mu$m emission line radiates as H$_2$ molecules cascade down to their lower states.
Infrared lines like 2.12$\mu$m penetrate through intervening dust, which absorbs ultraviolet
and optical light, and as such are ideal tracers of the interface between the ionized
and molecular gas when the ultraviolet radiation field is strong.}
\label{fig:1}
\end{figure}

Our H$_2$ observations were taken with the NOAO NEWFIRM camera on two observing runs,
20-22 Sept 2008 at the 4-m telescope at Kitt-Peak National Observatory in Arizona,
and at the 4-m telescope at the Cerro-Tololo Interamerican Observatory (CTIO) in Chile 16-18 March 2011. 
A single NEWFIRM image has a large field of view of $\sim$ 28 arcminutes, and a fine plate scale
of 0.4 arcseconds per pixel. The seeing ranged between 0.7 and 1.8 arcseconds over the course of the
observations with the exception of the H$_2$ and K-band images of M~16 (see Section 3.3).
The final images have a spatial resolution $\sim$ 4 times better than that possible with the
Spitzer Space Telescope. In addition, unlike Spitzer observations of H$_2$ that
rely upon broadband filters that transmit many emission lines as well as a large amount
of continuum, the NEWFIRM H$_2$ observations were taken through a narrowband filter
($\delta\lambda$/$\lambda$ $\sim$ 1\%) centered on the 1-0S(1) 2.12$\mu$m transition. As part of
the program we sometimes also obtained a narrowband 2.16 $\mu$m Br$\gamma$ image to detect
recombination from H in the H II region and to aid in continuum subtraction,
and a broadband 2.2$\mu$m K-image to identify
continuum from warm dust. A typical observing sequence involved a grid of up to 72 images dithered
spatially in a set pattern. Data reduction followed a procedure more fully-described elsewhere
\cite{carinaref}.

Infrared observations of the southern targets (Ara, Eagle and Carina) were complemented
by broadband optical B, V, R, and I and narrowband H$\alpha$ images taken with the
Y4KCam at the CTIO 1-m YALO telescope 12-17 March 2011. The Y4KCam is a 4064$\times$4064
backside-illuminated CCD with a plate scale of 0.289 arcseconds per pixel
and a field of view of 20 arcminutes.  Data reduction followed 
standard bias-subtraction and flatfielding procedures, with the additional
requirement that twilight sky flats were needed to remove a residual large-scale
illumination pattern. Several position dithers were used for each image to reduce internal
reflections from bright stars in the final median composite. We used a
distortion map determined from stellar positions to align the Y4KCam images with
the NEWFIRM data to within $\sim$ 0.1 arcseconds. All data reductions were performed within the IRAF environment. 

The narrowband H$_2$ filter blocks most of the continuum present in the K-band, but
all filters admit some continuum. It can be important to remove this component
when comparing the spatial positions of different emission lines along an interface,
and in regions where there are many background stars. However,
the continuum does not affect appearance of the PDR boundaries in the
images significantly because the H$_2$ bandpass is narrow enough that infrared continuum
contributions are small in the target objects. Subtracting a scaled K-band image to
remove continuum can be problematic
owing to slightly differing seeing conditions and sub-pixel alignment differences.  
For these reasons the narrowband images are uncorrected for continuum,
though we make use of the K-band to help interpret the regions, and in some cases
subtract the continuum with a scaled Br$\gamma$ or K-band image to create an image
that removes most stars and highlights the PDR interfaces.

\section{H$_2$ Images of PDR interfaces}

\subsection{Ara OB1}

Ara OB1 is a region of star formation in the southern sky centered $\sim$ 1.5 degrees out of the galactic plane
and subtending about a degree on the sky ($\sim$ 20 pc at its distance of 1320 pc, \cite{aradist}).
Its stellar population consists of a young open cluster, NGC 6193, whose most massive members are
an O7 star and an O5/O6 binary \cite{araostar}, similar to the situation in the Orion Nebula.
Ultraviolet radiation from these stars illuminates the Rim Nebula NGC 6188, a delicate undulating
feature located about 15 arcminutes (6 pc) to the west of the cluster. The Rim Nebula outlines
the edge of a dark cloud, within which lies an embedded star formation region known as RCW 108.
A CO survey of the region uncovered many distinct clouds along the line of sight \cite{araco}.

New H$\alpha$, Br$\gamma$, and H$_2$ images of the region are shown in Fig.~2. The H$\alpha$ image
highlights material as it boils off the PDR and recombines as it becomes ionized. Although the H$\alpha$
image outlines the interface between the ionized and neutral gas well, the boundary is not sharp and
is strongly affected by dust extinction. The multicolor composite in the middle panel shows a thin
red (H$_2$) feature which marks the actual boundary of the PDR where the UV light is absorbed. This boundary
is also clear in the difference image (Br$\gamma$ - H$_2$) in the right panel. The advantage of
the difference image is that Br$\gamma$ and H$_2$ 1-0S(1) have nearly the same wavelength, so
the relative amount of stellar continuum for each star is nearly the same for all sources, and as a result
the stars tend to subtract out. The H$_2$ feature circled in the right panel has a more clumpy
morphology and appears to originate from the RCW 108 region. This feature may be a stellar
jet that becomes visible as shock waves heat the ambient H$_2$. If it is a jet, the H$_2$ knots should exhibit
detectable proper motions on the sky after a suitable time interval.

Overall, the Rim Nebula is a wonderful example of an interface that appears more or less linear on the
sky, and exhibits the beginnings of pillar formation. Two features at the top of the H$_2$ image and one
at the bottom are $\sim$ 0.5 pc in size, with many smaller features present along the rim, though there is
a lack of detached irradiated globules. 
The ultraviolet radiation field in Ara OB1 is not particularly strong because the hottest star is only an O5, and
this results in a detectable, albeit moderate signal of H$_2$ from the PDR interface. As in all such
regions, without a dedicated numerical study one cannot readily determine if the undulating
structure present along the Rim Nebula results from variations in the gas density along the rim,
from instabilities induced by radiation, or some combination of the two.

\begin{figure}[b]
\includegraphics[width=5.0in]{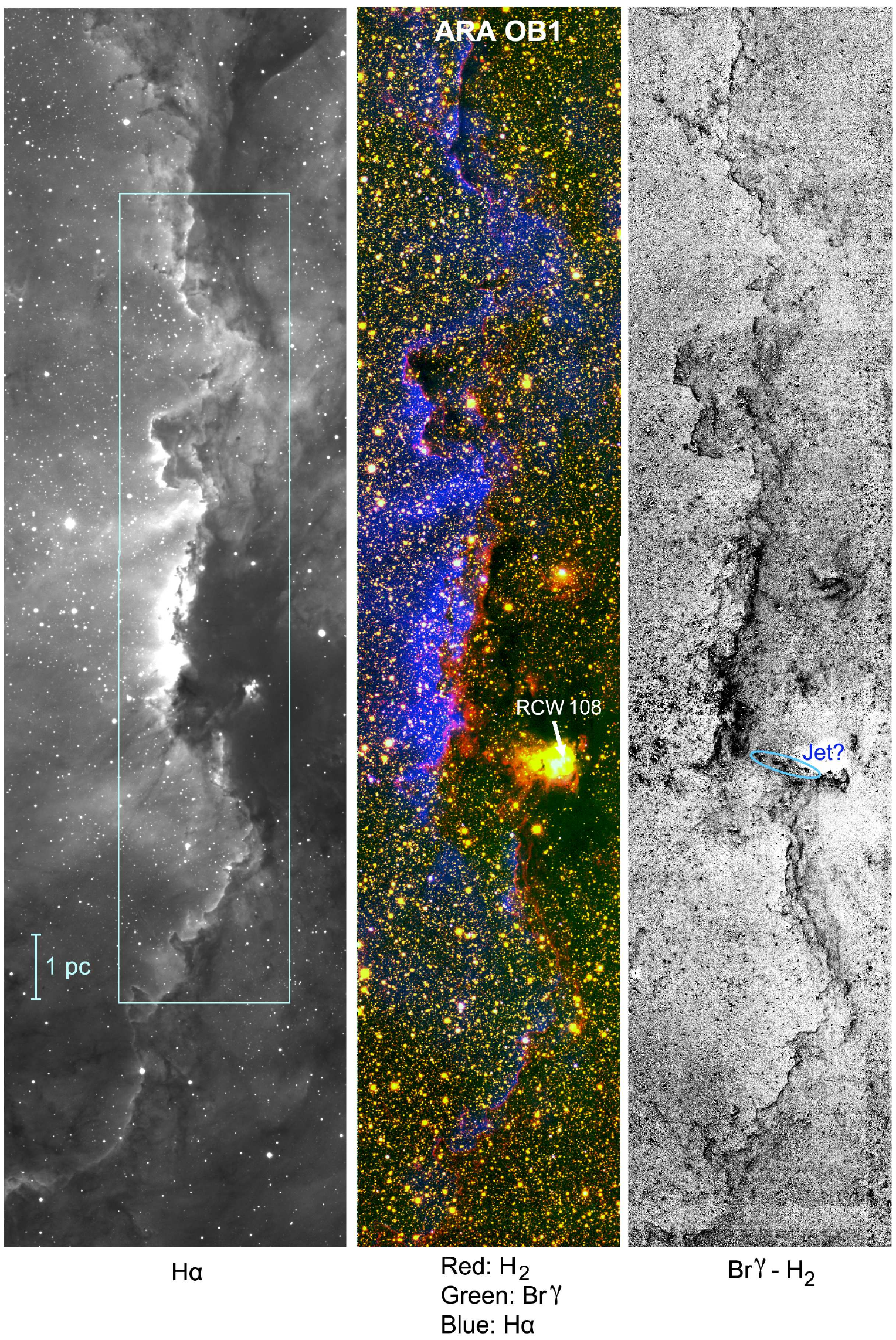}
\caption{Images of the Ara OB1 star formation region (standard convention of north up and east to the left).
Left: This narrowband H$\alpha$ image highlights recombining H gas as it boils off the dark cloud.
The cyan box shows the area covered in the center and right panels. The scale bar assumes a distance of 1320 pc.
Center: Color composite of the region showing the H$\alpha$ emission (blue) as it expands away from the 
PDR interface (H$_2$, shown in red). A large number of background and embedded stars appear yellow in the composite
because they are only present in the infrared (H$_2$, in red, and Br$\gamma$ in green) images.
The embedded star formation region RCW 108 is marked. Right: Subtraction image, where
H$_2$ emission appears black and Br$\gamma$ emission white.
Stars tend to subtract out in this image, which shows the boundary of the PDR in H$_2$.}
\label{fig:2}
\vfill\eject
\end{figure}

\subsection{Carina}

The star formation region in Carina is the premier example of its kind in the 
southern sky.  The region has $\sim$ 70 O stars,
more than any other OB association in the southern sky within 3.5 kpc, and excites a large
H~II region and nebula that is visible to the naked eye.
Its most famous resident is undoubtedly $\eta$ Car, a luminous blue variable (LBV)
that was once one of the brightest stars in the sky before it ejected a massive shell about
170 years ago\cite{eta}, and is a prime candidate for the next supernova in our galaxy.
Several clusters of massive stars, including Tr~14, Tr~15, Tr~16, Bo~10,
Bo~11, Coll~228, and Coll~232 exist in the vicinity.  The distance to the region is
uncertain owing to difficulties in accounting for reddening \cite{tr14}; a recent study found
2.9~kpc to both Tr~14 and Tr~16, \cite{carinadist}, so we will adopt this value. 
Carina is home to a number of O2 and O3 stars \cite{carinacensus}, and ages 
for the stellar content range from 0 to 5 Myr\cite{tr14}. 

The UV radiation field in Carina is dominated by two clusters. The first, Tr 16,
is a loose grouping that spreads out over $\sim$ 8 pc. It
contains 42 O stars in all, a WR star, $\eta$ Car, three O3 stars,
and shows evidence for subclustering \cite{carinaxray}.
The second cluster, Tr 14, is much more compact, and contains ten O stars, including an 
O2 star and an O3.5 star \cite{carinacensus,tr14}. Tr 16 is situated about a parsec
in projection south of $\eta$, some 10 pc to the southeast of Tr 14.
Optical images of the center of Car OB1 from HST and from the ground
provide a snapshot of the harsh, chaotic environment that surrounds a region
of massive star formation like Carina once the strong UV radiation and winds from
the nearby O and B stars begin to clear away the molecular clouds in their vicinity.
These images reveal a wealth of flows, globules, and star formation activity in the
region\cite{carhst}.

A full report of our Carina observations will appear elsewhere \cite{carinaref}, but
Fig.~3 provides a sample of the sort of PDR structures that are present in the region.
The color composite illustrates a striking transition between the highly-ionized [O III]
emission (blue) to the recombination lines of H (Br$\gamma$, green), to the fluoresced
H$_2$ line emission (red) as one proceeds into the molecular cloud. By magnifying the
bright PDR wall on the right side of the image, we see that the H$_2$ emission appears highly filamentary and
fragmented, and includes many ragged-shaped pillars.  Other large-scale wavy wall-like structures
are shown in panels B and G. Panel G overlaps with the top right corner of panel A, and
the cloud there is probably illuminated mainly by Tr 14. The object in panel B appears to have
a star near its apex, with a bright rim on its northwest side. Multiple exciting sources
may play a role here, as both Tr 16 and Tr 14 are located to the northwest, with
distances of 8.0 pc and 18.0 pc, respectively, to $\eta$ Car and to the center of Tr 14.

Panel C displays a small young cluster that has formed within a globule.
The globule is irradiated by sources to
the north, which may include stars in both Tr 16 (center $\sim$ 13 pc away) and Tr 14
(24 pc away), but the cluster has also created its own PDR along a cavity wall within the
globule on its northeast side. Spectacular irradiated pillars (E and F) and more complex
structures (D) are located in the same general area as the object in panel C and 
also respond to UV radiation from sources to the north. 

\begin{figure}[b]
\includegraphics[width=5.0in]{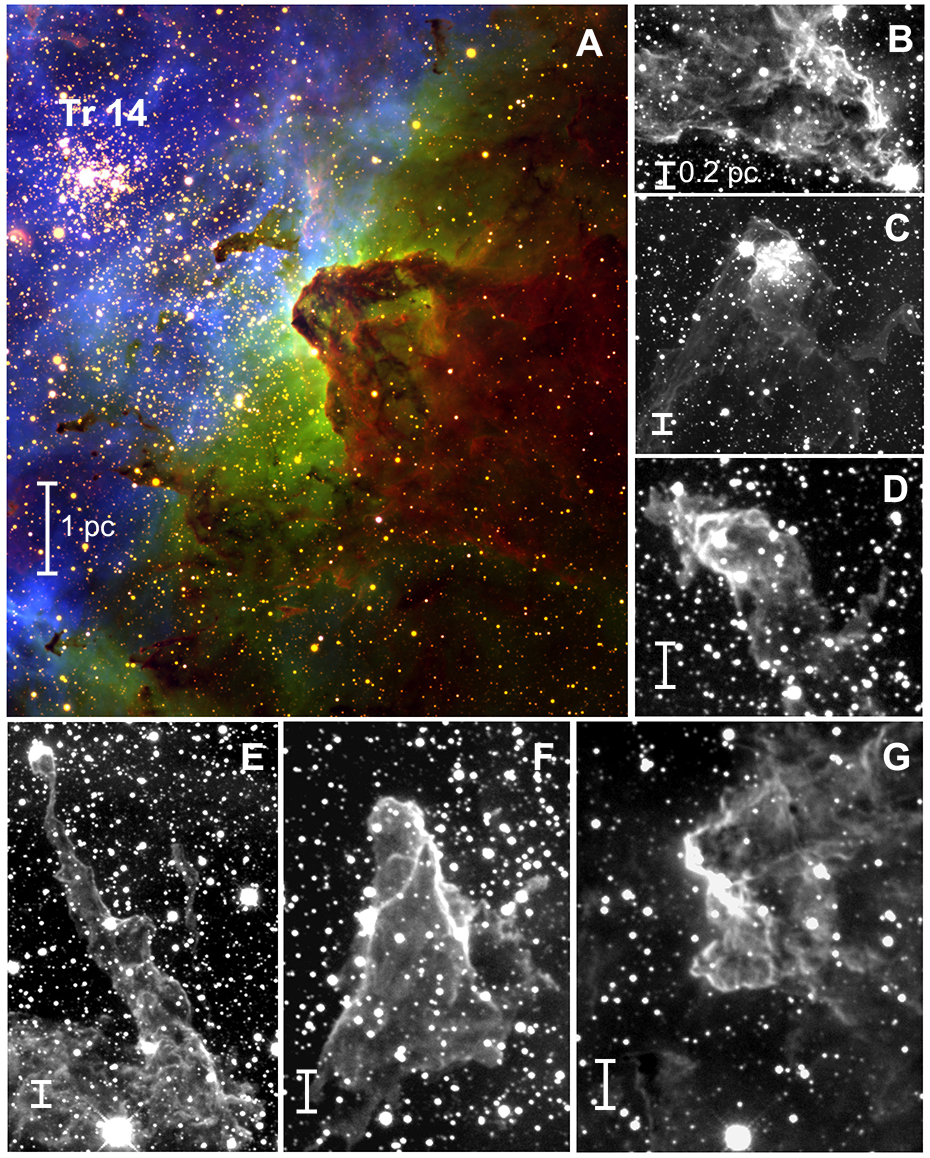}
\caption{Images of selected portions of the Carina star formation region.
The color composite has [O III] 5007\AA\ in blue, Br$\gamma$ in green and
H$_2$ 1-0S(1) in red. The compact cluster of O stars Tr 14 is marked.
The other panels are greyscale images of H$_2$ at various locations throughout the
region. Carina has spectacular pillars, walls, and proplyds.
Scale bars assume a distance of 2.9 kpc, and are 0.2 pc in all panels 
except for the color composite. Each panel is discussed in the text.
}
\label{fig:3}
\end{figure}

It is clear that Carina is a very rich environment where one can study how molecular
clouds transform when they are subjected to intense UV radiation fields and strong stellar winds.
Structures of all types abound here, including many varieties of walls, pillars, and
detached globules. In response to the UV fluxes, the fluoresced H$_2$ emission is quite bright, 
even along the sides of the pillars.  One challenge with interpreting
the region is that many possible ionizing sources exist and the radiation does not necessarily
mainly come from a single direction. 

\subsection{Eagle Nebula}

The Eagle Nebula (M 16) is instantly recognizable to the public because of the 
so-called 'Pillars of Creation' images published during the early days of the
Hubble Space Telescope \cite{m16poc}, and subsequent images of the 'Spire' taken as
part of the Hubble Heritage project.  The Pillars have been studied in molecular line
emission \cite{m16co}, and recent space-based surveys include 
Spitzer observations in the near-infrared \cite{m16spitz} and
Herschel observations in the mid- and far-infrared \cite{m16her}.
In a study of the Pillars with a Fabry-Perot, Allen et al.
observed H$_2$ near the apices of the Pillars, and found this emission to be offset from
Br$\gamma$ in the direction away from the ionizing sources \cite{m16fluor}.  The sharpest images 
of the Pillars in the H$_2$ and Br$\gamma$ lines come from the NICMOS camera on HST \cite{m16nicmos},
and show this spatial offset clearly, along with exquisite detail of the H$_2$ that resembles
the structured filaments we observe in Carina (Fig.~3).
However, the very small field of view of NICMOS limited these
observations to the immediate area around the heads of the Pillars.

Using a distance to the Eagle of 1.8 kpc \cite{m16dist},
the Pillars and the Spire are located $\sim$ 2 pc to the southeast and east, respectively, of
a dense open cluster of stars known as NGC 6611. This cluster contains 
about a dozen O stars, with the most massive being an O4
\cite{m16starcensus}.  NGC 6611 has ejected several lower-mass O stars that form bow shocks as
they move through the interstellar medium \cite{m16eject}.
The M~16 area has ongoing star formation, with typical ages
of a few million years for the young stars \cite{trigger, m16age}. 
North of the cluster, the Herschel and Spitzer
images show what appears to be the ridge of a dark cloud, and a fainter arch that
extends further to the north \cite{m16spitz, m16her}.
Based on the count of O-stars, the Eagle Nebula
falls somewhere between Ara OB1 and Carina on the scale of massive star formation.

Our observations of the Eagle appear in Fig.~4. The H$2$ images were taken in
poor seeing conditions (2.8 arcseconds, as compared to 0.8 arcseconds for Br$\gamma$) which has the
effect of blurring out sharp features. Nonetheless one can identify that the main interaction
regions are the Pillars, the Spire, and the Ridge, with the most extensive H$_2$ located in the
Ridge. Spatial offsets between the Br$\gamma$ and H$_2$ are in the same sense that we observed
in Ara and Carina, with the Br$\gamma$ emission located closer to the ionizing sources in the cluster.
The cluster appears particularly dense in the infrared (lower left portion of the Ridge image).
At the southern edge of the North Bay, a dark feature visible at optical wavelengths, we observe 
knots of H$_2$ emission that may represent a molecular jet. The Spitzer
IRAC image at 8 $\mu$m also shows this feature, albeit with rather poor spatial resolution
\cite{m16spitz}.

\begin{figure}[b]
\includegraphics[width=5.0in]{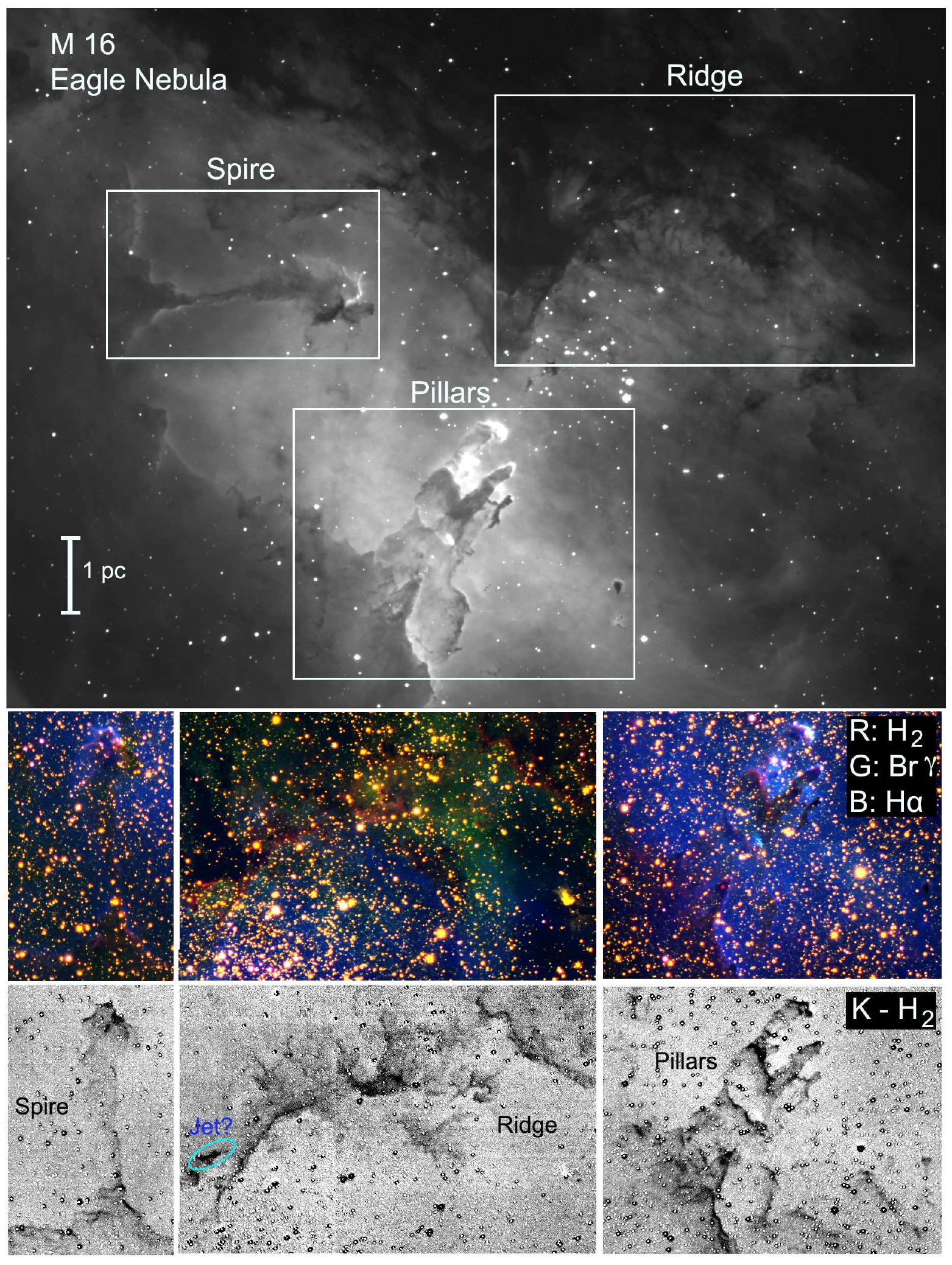}
\caption{Images of the Eagle Nebula. Top: H$\alpha$ image, showing the locations of the Spire, Ridge, and Pillars
that are magnified in the subsequent panels. The North Bay is the dark region along the left of the Ridge
image. Middle: Color composites of the magnified regions, with H$\alpha$ in blue,
Br$\gamma$ in green, and H$_2$ 1-0S(1) in red. 
Bottom: Difference between a continuum band (K) and a narrowband H$_2$ image, highlighting the PDR boundaries.
The scale bar assumes a distance of 1.8 kpc. 
}
\label{fig:4}
\end{figure}

Despite the notoriety of its famous pillars, the Eagle Nebula is only a modest hunting
ground for irradiated interfaces. As Thompson et al. note \cite{m16nicmos}, much of the star formation activity
may be finished in the Eagle, leaving only a few areas like the Ridge and the Pillars for 
irradiated interfaces.  Apart from the possible jet at the bottom of the North Bay, our
H$_2$ images do not reveal any striking new features, though in addition to the well-known Pillars and Spire,
we can identify the Ridge as a promising area where extensive, and relatively bright PDR interfaces exist.
Images of the Ridge taken during better seeing conditions would likely resolve the H$_2$ emission
into a filamentary, bumpy wall like we observed in Ara OB1.

\subsection{Cygnus OB2}

If one were to fly far above the galactic plane and look down upon the
structure of the Milky Way within a few kpc of the Sun, a single star
formation region would stand out as being the major area of activity
in this portion of the galaxy: Cyg OB2. Also known as Cygnus X,
the region houses over 120 O stars, including several of type O3 \cite{cygostar},
has a candidate for the most massive star in the galaxy,
evidence for recent supernovae \cite{cygsnr}, and may even be the source of
a large-scale galactic fountain \cite{cygbubble}. The region would be
a naked-eye cluster known to all amateur astronomers were it not for the
fact that it lies near the galactic plane along the Cygnus arm, and
for that reason suffers 10 magnitudes of visual extinction along the line of
sight. X-ray observations reveal at least a thousand young stars
in the region \cite{cygxray}.

The new H$_2$ observations cover an area over 1.5 degrees on a side, and include
the entire star-forming region. Fig.~5 depicts some of the more interesting
of these structures. A dense cluster of massive stars anchored by the O3 star
S 22 at the core of the region (panel D) is the main source of radiation, although
O stars are scattered throughout the area.  Surrounding the core we find dozens of
irradiated globules that possess head-tail structures with the tails pointing
away from the ionizing sources.  Several of these are shown in Fig.~5, along with
an arrow that denotes the distance to S 22.  Some of these objects have been 
discovered independently by other authors and described in recent papers
\cite{cygglob1,cygglob2}. 

Each panel demonstrates an important aspect of radiation hydrodynamics in 
the star formation process. Panel A shows that two stars within this globule are
surrounded by reflection nebulae, indicating that stars still form here. Narrow
filaments of H$_2$ are associated with the stars in the globule as they create
their own cavity from an outflowing wind, but the overall shape of the PDR is
still driven by the cluster of O stars at the large projected distance of
over 20 pc. Hence, young massive stars like those in Cyg OB2 can influence 
the ongoing star formation over a very large volume. This motif repeats in panel C,
where a young star at the apex of the globule appears to have created enough of
a cavity to nearly detach from its globule. Panel B depicts what appears to be an
earlier stage reminiscent of the wall in Ara OB1, where small pillars are beginning
to form.

Panels E and F display small remnants of globules
that appear similar to objects known as proplyds in the Orion Nebula. The proplyd 
within panel E possesses a wonderful `wiggling' tail that resembles a Kelvin-Helmholtz
instability (see also \cite{cygglob1}). However, it is unclear what role
stellar winds from the core play in creating this shape, as the wind should first 
encounter a strong bow shock around the obstacle before the postshock wind could
interact with the neutral material to create a shear layer that would induce mass
stripping. The H$_2$ images here show only the boundary where the molecular gas
becomes excited by direct UV radiation from the core stars and by ambient UV
radiation scattered within the H II region or originating from less massive O stars
in the vicinity. It is likely that a bow shock is present ahead of the proplyd,
but this shock would be difficult to detect if the wind from the massive stars is fully ionized. 
The wave-like structure along the sides of the proplyd in panel E
has a characteristic length on the order of the size
of the dense clump at the head of the globule.
Panel F shows similar structures, but not all of these have formed a star at their apices.

\begin{figure}[b]
\includegraphics[width=5.0in]{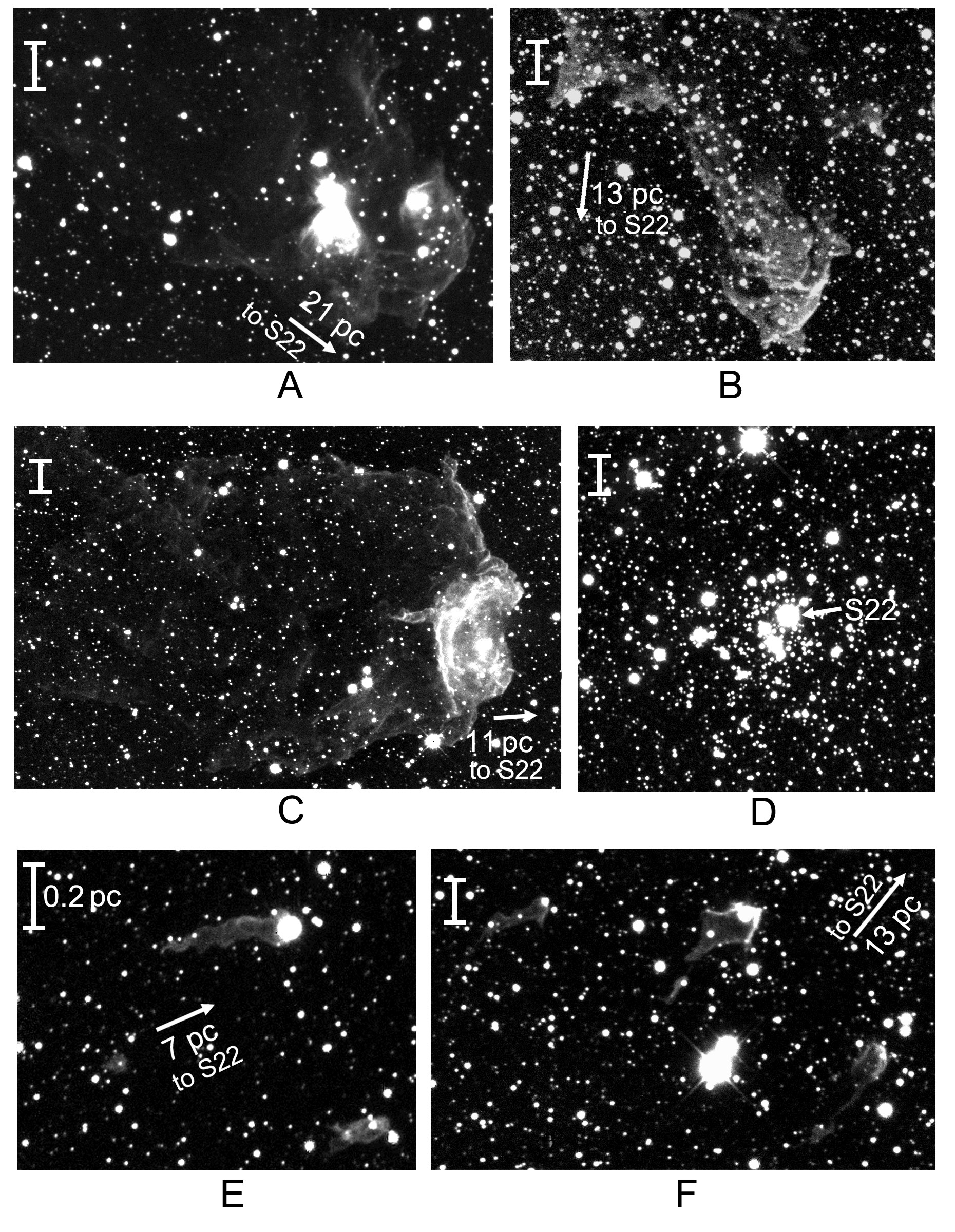}
\caption{H$_2$ images of globules in the Cyg OB2 star formation region.
The brightest O star in the central cluster, Schulte 22, is labeled in panel D
and forms the nominal center of the region. Arrows in each panel indicate the
distance (from the base of the arrow) and direction to S 22.
Objects within each of the panels are discussed
in the text.  North is up and east to the left, and the scale bars are 0.2 pc 
for each panel, and assume a distance of 1.7kpc to the system.}
\label{fig:5}
\end{figure}

The Cyg OB2 region is among the most massive star formation regions in our galaxy.
The many O stars scattered throughout the region and the lack of bright irradiated walls
near the cluster suggests may be several million years
older than our other targets \cite{cygage}, though some star formation
continues within the few remaining globules. These globules, probably the densest
leftovers from the original molecular cloud, glow along their boundaries
in H$_2$ as they respond to the intense UV ambient radiation field and possibly strong
stellar winds from O stars in the area. In some cases 
radiation from one or more young stars at the head of the globule also affects the
dynamics of the system.  In Cyg OB2 we are witnessing the last
stages of the photoablation processes in a region where the most massive stars have
already formed and have cleared away most of the ambient material.

\section{Summary}

Regions of massive star formation offer some of the best and most easily-observed
examples of radiation hydrodynamics. The boundary where the fully ionized H~II
region transitions to the neutral/molecular photodissociation region is of
particular interest because the marked temperature and density contrasts across
the boundary lead to evaporative flows and fluid dynamical instabilities that can
develop into spectacular pillar-like structures. Pillars subsequently become ionized
globules when they detach from their parent clouds, and the globules often harbor
one or more young stars at their apices. It is possible to observe the interface between the H~II
region and PDR directly by imaging in the light of the H$_2$ 1-0S(1) transition, which
occurs as a byproduct of fluorescent excitation of H$_2$ to the Lyman and Werner bands.
Extinction by dust is negligible at 2.12$\mu$m, so it is possible to observe this interface
directly, and with better spatial resolution than was possible from wide-field space-based
missions such as Spitzer and Herschel.

The new images illustrate how radiation can sculpt molecular clouds into a variety of spectacular
forms. Notwithstanding the dizzying array of morphologies,
there appears to be a progression from (i) the wavy wall-like structure in Ara OB1, where
radiation appears to have had as yet only a modest influence on the molecular cloud, to (ii) the
intricate pillars and cavity walls within Carina and to a lesser extent in the Eagle Nebula,
where radiation and outflows are in the process of tearing the cloud apart, to (iii) the relatively isolated
head-tail globules in Cygnus OB2, where the massive stars have already dissipated much of
the natal molecular cloud, and leave behind only small remnants.

Future efforts by our group will focus on quantifying how spatial offsets between different emission
lines in the PDR change as the ultraviolet radiation source varies in strength. Finding evidence for
or against stellar wind shocks is another important observational goal, as the massive stars which 
generate the UV radiation field can also drive powerful winds that disrupt their surroundings.
Numerically, there is a clear challenge to predict offsets between different emission lines in
a PDR, and to distinguish irregular interfaces where the shape is driven by radiative instabilities
from those produced by dense clumps within the molecular cloud. Combining radiation with winds
should be another numerical frontier. On the experimental side,
new programs at Omega and NIF should begin to explore the regime of radiative shocks and strongly-irradiated
interfaces \cite{nifref1,nifref2,nifref3}, and can look for similar structures to the ones presented here.

\section{Acknowledgements}

This work was made possible through DOE funding via the NLUF program. We thank R. Probst for his
assistance with the KPNO observations, the staff of CTIO for their support at the 4-m and 1-m telescopes
and W. Henney for useful discussions about PDRs. 








\end{document}